# Improving electrical properties of ferroelectric thin films by ultraviolet radiation

Neil McN. Alford [1], Aleksandr G. Gagarin [2], Andrey B. Kozyrev [2], Peter Kr. Petrov [1], Aleksandr I. Sokolov [2], Oleg I. Soldatenkov [2], and Valery A. Volpyas [2]

[1] *Center for Physical Electronics and Materials, Faculty of Engineering, Science and the Build Environment, London South Bank University, 103 Borough Road, London SE1 0AA, UK*

[2] *Saint Petersburg Electrotechnical University, 5 Professor Popov Street, St. Petersburg, 197376, Russia*

*Abstract*

Ferroelectric films in the paraelectric phase exhibit two undesirable properties: hysteresis in the voltage-capacitance characteristics and a significant relaxation time of the capacitance. Our experiments show that suppression of both of these is achieved by using ultraviolet (UV) radiation with wavelengths corresponding to the material forbidden gap. Experimentally we also observed UV radiation induced modulation of thin film permittivity without an applied electric field. The observed phenomena are believed to have the same origin: UV light causes generation of non-equilibrium charge carriers and space charge redistribution inside thin ferroelectric films.

Ferroelectric materials have been the subject of extensive research for over 50 years. However, recent investigations of thin films of SrTiO$_3$ (STO), BaTiO$_3$ (BTO), and their solid solution Ba$_x$Sr$_{1-x}$TiO$_3$ (BSTO), motivated by the prospect of new applications such as electrically controllable microwave devices[1,2], have uncovered many complexities not previously recognised. Most effort has been aimed at the optimisation of thin film fabrication processes, i. e. to the microwave loss reduction (microwave loss tangent of ferroelectric thin films is much higher than that of corresponding bulk crystals)[3,4], increasing the tunability[5], and improving the temperature stability[6]. These points are now better understood, but the issue of the residual polarization and the hysteresis phenomenon observed under varying bias voltage, resulting in slow relaxation of dielectric constant of ferroelectric films in paraelectric phase (above the Curie temperature)[7], has not properly been addressed.

In principle, ferroelectric materials in the paraelectric phase should not exhibit hysteresis in the C(V) behaviour (voltage–capacitance characteristics)[8]. They should demonstrate also very short response time comparable with the period of the lattice soft mode oscillations. However, to the best of our knowledge, there are still no hysteresis-less devices or devices with switching time close to the theoretically predicted few picoseconds, reported in the literature. Hysteresis and related phenomena result in a decrease of tunability and an ambiguous response time of the ferroelectric materials. Overcoming this problem will enable broader application of the ferroelectrics in microwave electronics.

There exist a number of models explaining the residual polarization and dielectric relaxation in ferroelectric thin films. The charge injection[9] and space charge formation[10] models, the Debye-type relaxation model[11], and the hopping probabilities distribution model[12] are among them. However, the real physical mechanisms of the discussed phenomena require further clarification.

In this letter we present a method, validated by experimental results, to solve the hysteresis problem in ferroelectric thin films using UV light radiation.

Previous works examining the application of UV radiation to ferroelectrics considered mainly bulk ferroelectric materials and were devoted to the photo-controllable phase transitions and the UV light induced shift of the Curie temperature[13-16]. They were motivated by the fact that the contribution from the electronic subsystem (excited atom states, non-equilibrium charge carriers, etc.) to the free energy of the crystal might be substantial. The photo-ferroelectric effects in bulk ferroelectrics with no internal electrical field are caused by the optical excitation of the non-equilibrium charge carriers.

In our investigation, we irradiated thin BSTO films in the paraelectric phase by UV light with wavelengths between 350 nm and 430 nm. UV light with such wave lengths can generate excess charge carriers in the films since the corresponding photon energies (2.9-3.7 eV) are close to the forbidden gap of BSTO material (3.3-3.8 eV)[17]. In turn, the excess charge carriers can affect the electrical properties of the films by suppressing the non-uniform charge distribution and screening charged defects.

The structures investigated were planar capacitors formed on $Ba_{0.3}Sr_{0.7}TiO_3$ films deposited by RF magnetron sputtering and laser ablation on $Al_2O_3$ and $LaAlO_3$ substrates, respectively. Details concerning the deposition processes can be found elsewhere[18,19]. Various samples with thickness between 0.5 μm and 1 μm were investigated. The Curie temperature of all samples was measured to be about 140 K, which is in a good agreement with that measured for the bulk materials. The estimated dielectric permittivity (ε') at zero d. c. voltage and room temperature varied between 300 and 500. The electrodes (bi-layers of Au/Ti or Cu/Cr) were formed by thermal evaporation followed by lift-off or wet-etching, respectively. The metal layer thickness was between 0.5 μm and 1 μm, while the gap between the electrodes varied from 5 μm to 20 μm.

The sample capacitance was measured at room temperature (300 K) under reverse d.c. bias in 0 – 300 V range, using LRC meters E7-12 (1 MHz) and Agilent 4287A (1 GHz). The error in both measurements did not exceed few femtoFarads. The results obtained were similar for all samples at radio frequencies (RF) and microwaves, therefore experimental data presented below are in arbitrary units and measurement frequency is not mentioned.

Three types of experiment were performed:
1. Voltage-capacitance characteristic of the capacitors were measured with and without UV irradiation. The UV light source was a GaAs light emitting diode (LED) with wavelength 370 nm and power density 7.5 mW/cm$^2$. Typical results are presented in Fig. 1. Curves 1 and 2 show the reversible change of the sample capacitance under d. c. bias changed from 0 V to 300 V and back to 0 V without UV irradiation. As one can see, a well-pronounced hysteresis of the capacitance takes place. Under UV light, a slight increase of the capacitance is observed at 0 V (see curve 3), which results in steeper behaviour of the voltage-capacitance characteristics at low d. c. bias. At high d. c. bias (close to the $U_{max}$), however, the characteristics become identical. On decreasing voltage back to zero, the measured capacitance values are identical with those previously measured on the voltage increase, i. e. no hysteresis is observed. Similar results were obtained for all measured samples irradiated with UV light with wavelength between 350 nm and 430 nm. At longer wavelengths, however, the removal of hysteresis was not complete: it was reduced substantially in comparison with the case without UV irradiation, but was still detectable.

A possible explanation of the described phenomenon (hysteresis disappearance) is that the UV photons with energy close to the BSTO band gap (3.3 – 3.8 eV) generate non-equilibrium charge carriers, which neutralize the space charge trapped by the film defects. The incomplete suppression of the hysteresis in some of the samples might be due to insufficient (too low) energy of the UV photons with longer wavelengths. It should be noted that the capacitor structures were not optimised for these measurements.

2. Measurements were made to investigate the influence of the UV light irradiation on the capacitance/permittivity relaxation time after the end of control voltage pulse. Rectangular pulses with amplitudes up to 300 V and duration between 10 s and 30 s were applied to the same capacitor structures. The capacitance was measured with and without UV light. Typical results are presented in Fig. 2. Curves 1 and 2 represent the capacitor response to a rectangular pulse with 10 s duration showing conventional relaxation type behaviour. Without UV light, the capacitance/permittivity relaxation achieved 90% of the initial value after several tens of seconds (curve 1, Fig. 2). With UV irradiation the relaxation time was significantly reduced (curve 2, Fig. 2), approaching the theoretical limit for the ferroelectric materials.

3. In both experiments, all samples under UV irradiation and 0 V bias exhibited capacitance higher than that measured in the virgin state. This led us to the idea of the third experiment: modulation of the BSTO capacitance by an irradiation of modulated ultraviolet light. The results are presented in Fig. 3. As one can see, the change in capacitance follows exactly the time modulated (0.5 Hz) UV radiation. Again, although the experimental set up used was unable to measure the relaxation time, it was well below a few milliseconds (the equipment refreshment time). We believe that the origin of the observed phenomena is the photo-induced generation of non-equilibrium charge carriers which reduces a local electric field. This affects the local dielectric permittivity and changes the effective permittivity of the sample. The change in absolute capacitance value was about 3%. However, as mentioned above, the capacitor design was not optimized for these measurements.

In principle, the variation of sample capacitance/permittivity under UV radiation may be also due to its heating. To exclude this possibility, we estimated the UV radiation induced variation of the sample temperature. Taking into account that BSTO dielectric permittivity $\varepsilon$' at ~ 800 THz (λ ~ 370nm) is of order of 10, the UV induced change in sample temperature was found to be about $0.01^0$, which is far below the value that could warrant the measured change in sample capacitance.

In conclusion, our investigation shows that using UV radiation with wavelengths corresponding to the BSTO forbidden gap it is possible to remove or significantly suppress the hysteresis in voltage-capacitance characteristics of the BSTO based nonlinear capacitors and to strongly reduce their capacitance/permittivity relaxation time. Experimentally we observed that BSTO thin film permittivity is sensitive to the UV radiation even without an applied electric field. The observed phenomena are considered to have the same origin. Namely, UV light causes generation of non-equilibrium charge carriers that screen out local electric field inside the BSTO thin films and change their effective dielectric properties.

This work was partly supported by the Engineering and Physical Sciences Research Consul (EPSRC), UK. A.I.S. acknowledges the financial support of the Russian Foundation for Basic Research under Grant No. 04-02-16189.

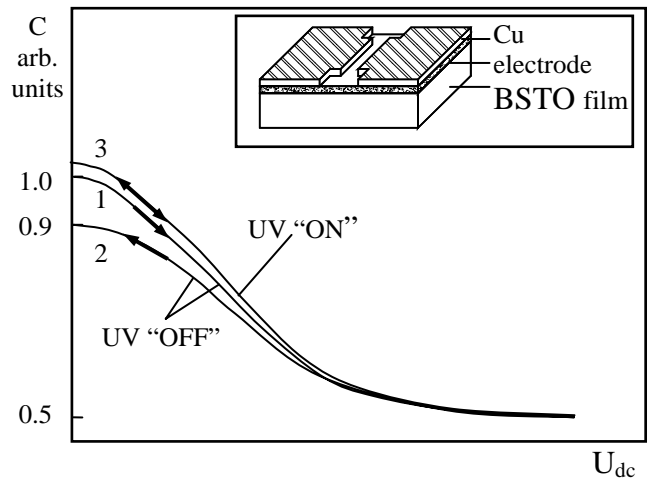

**Figure 1.** Capacitance of the BSTO capacitor versus dc bias voltage at absence of ultraviolet exposure (curves 1 and 2 refer to the increase and decrease of bias voltage, respectively) and at ultraviolet exposure (curve 3). Insert shows the structure of ferroelectric film planar capacitor investigated.

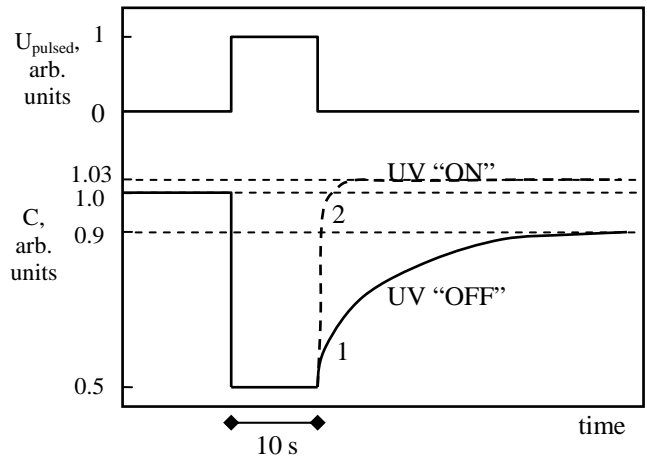

**Figure 2.** Evolution in time of the capacitance of the BSTO capacitor after the pulsed control voltage is switched off without (curve 1) and under (curve 2) the exposure of ultraviolet light.

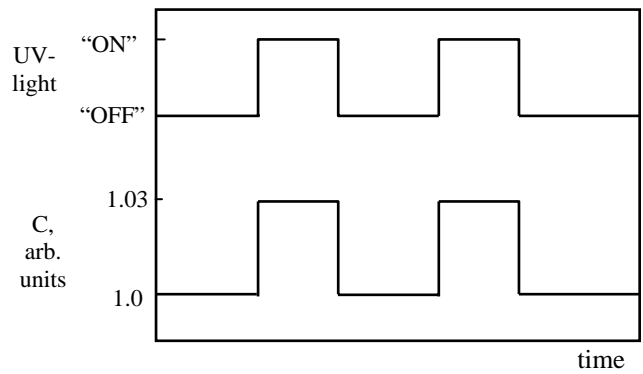

**Figure 3.** The modulation of the BSTO capacitance due to modulated ultraviolet light. No bias voltage was applied.

7